\documentclass[aps,pre,floatfix,preprint,superscriptaddress,nofootinbib,a4paper]{revtex4-1}
\usepackage{amsmath,amssymb}
\usepackage{graphicx}
\usepackage{dcolumn}
\usepackage{bm}
\usepackage{subfig}
\usepackage{paralist}
\usepackage{color}

\def\<{\langle}
\def\>{\rangle}
\def\(({\left(}
\def\)){\right)}
\def\[[{\left[}
\def\]]{\right]}

\newcommand{\sDr}{$(d,\sigma) \leftrightarrow D$ }

\begin{document}

\title{Relations between Short Range and Long Range Ising models}

\author{Maria Chiara Angelini}
\affiliation{Institut de Physique Th\'{e}orique, CEA/DSM/IPhT-CNRS/URA 2306 CEA-Saclay, F-91191 Gif-sur-Yvette, France}

\author{Giorgio Parisi}
\affiliation{Dipartimento di Fisica, INFN -- Sezione di Roma 1, CNR -- IPCF UOS Roma, Universit\`{a} ``La Sapienza'', P.le A. Moro 5, I-00185 Roma, Italy}

\author{Federico Ricci-Tersenghi}
\affiliation{Dipartimento di Fisica, INFN -- Sezione di Roma 1, CNR -- IPCF UOS Roma, Universit\`{a} ``La Sapienza'', P.le A. Moro 5, I-00185 Roma, Italy}

\begin{abstract}
We perform a numerical study of the long range (LR) ferromagnetic Ising model with power law decaying interactions ($J \propto r^{-d-\sigma}$) both on a one-dimensional chain ($d=1$) and on a square lattice ($d=2$). We use advanced cluster algorithms to avoid the critical slowing down. We first check the validity of the relation connecting the critical behavior of the LR model with parameters $(d,\sigma)$ to that of a short range (SR) model in an equivalent dimension $D$. We then study the critical behavior of the $d=2$ LR model close to the lower critical $\sigma$, uncovering that the spatial correlation function decays with two different power laws: the effect of the subdominant power law is much stronger than finite size effects and actually makes the estimate of critical exponents very subtle. By including this subdominant power law, the numerical data are consistent with the standard renormalization group (RG) prediction by Sak, thus making not necessary (and unlikely, according to Occam's razor)  the 
recent proposal by Picco of having a new set of RG fixed points, in addition to the mean-field one and the SR one.
\end{abstract}

\maketitle

\baselineskip = 20pt

\section{Introduction}

It is well known that ferromagnetic (FM) systems of discrete spins with a finite range of interaction have a lower critical dimension $D_L=1$. 
It means that a one-dimensional chain of spins can not undergo a phase transition at any positive temperature \cite{D1SR}.
The situation is different if one considers long-range (LR) models \cite{Dyson} in $d$ dimensions.
They are fully connected models, with a Hamiltonian:
\begin{equation}\label{Eq:H}
 H=-\frac{1}{2}\sum_{i,j=1}^{N}J_{ij}\sigma_i\sigma_j\;.
\end{equation}
The range of interactions is infinite and the intensity of the coupling $J_{ij}$  
decays as a power law with the distance between spins: $J_{ij}\propto|r_{ij}|^{-(d+\sigma)}$. 
One can also define spin glasses on LR models, taking $J_{ij}$ as independent identically distributed random variables,
extracted from a distribution $P(J)$ (like for example a binary or Gaussian distribution),
requiring that the variance of $P(J)$ decays as a power law:
$\overline{J_{ij}^2}\propto |r_{ij}|^{-(d+\sigma)}$ \cite{LR-SG}.
These models can have a transition at dimensions smaller than the lower critical one for usual short range (SR) models. 
Indeed a ferromagnetic LR model can have a transition also in $d=1$. Furthermore, varying $\sigma$, 
the behaviour of the system (such as for example the critical exponents) can vary from a mean field to a non mean field one 
until it reaches a certain value $\sigma_L$ and the behaviour of the corresponding SR system is recovered. 
In fact one can write down a relation between $(d,\sigma)$ and the effective dimension $D$ of an equivalent SR model. 
At $\sigma_L$, the effective dimension $D$ reduces to the real dimension $d$ of the LR system. 
The behaviour in and out the range of validity of mean field approximation can thus be observed varying only a parameter, 
and this is very useful if one wants to simulate the system numerically because the computational complexity of the model does not change 
with the effective dimension.
Different LR models have been introduced in the past, and more than one relation \sDr exist. 
Nonetheless, are still unclear the differences between various LR models, 
the exactness of the \sDr relations and their limits.
Often LR models (both ferromagnetic and disordered ones) 
have been used to extract properties of the analogous SR effective models, however it is not clear whether 
this operation is really justified.

The purpose of this work is to summarize the previous works on LR models, 
and to answer some crucial questions on how good are LR models to simulate SR models,  
which is the best \sDr relation, what is its range of validity and how similar are different LR models. 
Many of the answers are unknown even in the simplest ferromangetic case, for this reason we will mostly analyze FM systems.

\section{Review of known analytical and numerical results for ferromagnetic Long Range models}

The simplest LR model that can undergo a paramagnetic/ferromagnetic phase transition is a one-dimensional 
chain of spins with the Hamiltonian of Eq.~(\ref{Eq:H}) and $J_{ij}\propto |i-j|^{-(1+\sigma)}$.
For this model, Dyson demonstrated analytically that there is a standard second order phase transition if $0<\sigma<1$ \cite{Dyson}. 
For $\sigma\leq 0$ the energy is no longer an extensive quantity.

This model can be easily generalized to $d$ dimensions, redefining the couplings as $J_{ij}\propto r_{ij}^{-(d+\sigma)}$ where $r_{ij}$ is the euclidean distance $r_{ij}=|\vec{r}_i-\vec{r}_j|$.

This model has been analyzed using a renormalization group (RG) approach in Ref.~\cite{fieldtheoryLR}.
The field theory in the momentum space can be written as:

\begin{equation}
\begin{aligned}
 \int dx \mathcal{L}(\phi)= \sum_ku_2(k)\phi(k)\phi(-k)+u\sum_{k_1 k_2 k_3}\phi(k_1)\phi(k_2)\phi(k_3)\phi(-k_1-k_2-k_3)
\label{Eq:LRLagrangian}
\end{aligned}
\end{equation}
where $u_2(k)=r+j_{\sigma}k^{\sigma}+j_2k^2$, and the parameter $r$ varies linearly with the temperature, being null at criticality.

The intuitive, but too naive, interpretation given in Ref.~\cite{fieldtheoryLR} is the following:
if $\sigma>2$, the leading term in $u_2(k)$ is the $k^2$ one, then the usual SR behaviour in $d$ dimensions is recovered; while, for $\sigma<2$, the leading term in $u_2(k)$ is the $k^{\sigma}$ one, the $k^2$ term being subleading, and a behaviour different from the SR one is present.

If $2\sigma-d<0$ the Gaussian fixed point $u^*=r^*=0$ is stable.
The critical exponents are easily calculated, leading to $\nu=1/\sigma$, $\eta=2-\sigma$, $\gamma=1$.
At $\epsilon=2\sigma-d=0$, this fixed point is marginally stable and logarithmic behaviour appears for the correlation length and susceptibility.
This point corresponds to the upper critical exponent $\sigma_U=\frac{d}{2}$.
For $\epsilon>0$ the Gaussian fixed point is unstable with respect to $u$ and a new fixed point $u^*=O(\epsilon)$ is found.
The critical exponents can be obtained as a series expansion in $\epsilon$.
$\eta$ is found to be not renormalized up to third order in $\epsilon$ and it is commonly believed that
it will have the mean-field value at all orders because new $k^{\sigma}\phi(k)\phi(-k)$ terms are not generated under renormalization. 
This has been also verified numerically with good accuracy in Ref.~\cite{LR_SRnumeric}.

Summarizing, the picture that emerges from the work of Ref.~\cite{fieldtheoryLR} is the following:
for $0<\sigma<\sigma_U=\frac{d}{2}$ the system is in a mean-field region, for $\sigma_U<\sigma<\sigma_L=2$ 
the exponents are different from the mean field ones and
change continuously with $\sigma$, for $\sigma>\sigma_L$ the SR behaviour is recovered.
However there are some debated points.
For example, in this picture the lower critical exponent is $\sigma_L=2$, for all the dimensions\footnote{\baselineskip 15pt
In SR systems the lower critical dimension is the dimension at which the phase transition ceases to exist. 
In LR models, we call lower critical exponent the value of $\sigma$ such that the SR behaviour is recovered.
If the SR model has a phase transition in $d$ dimensions, then there is a transition also for $\sigma>\sigma_L$; on the contrary, as in $d=1$, there is no transition for $\sigma>\sigma_L$.}.
However in $d=1$, at $\sigma=1$ the transition becomes of the Kosterlitz-Thouless (KT) type \cite{KT}, 
supported by analytical \cite{LCD-analytical} \cite{LCD-analytical2} and numerical \cite{LCD-numerical} evidences.

The problem of the inconsistency of the results of \cite{fieldtheoryLR} near $\sigma=2$ 
is not related only to the one-dimensional case.
In fact, according to the picture of Ref.~\cite{fieldtheoryLR}, $\eta=2-\sigma$ for $\sigma<2$ and $\eta=\eta_{SR}$, for $\sigma>2$.
This would imply a jump discontinuity in $\eta$ at $\sigma = 2$, and a non-monotonic behaviour in $\sigma$. While this
phenomenon is not forbidden by thermodynamic arguments (which only require $\eta\leq2+\sigma$), it has
attracted considerable attention over the past decades, because it is quite singular. 

\begin{figure}[t]
\begin{center}
\includegraphics[width=0.7\columnwidth]{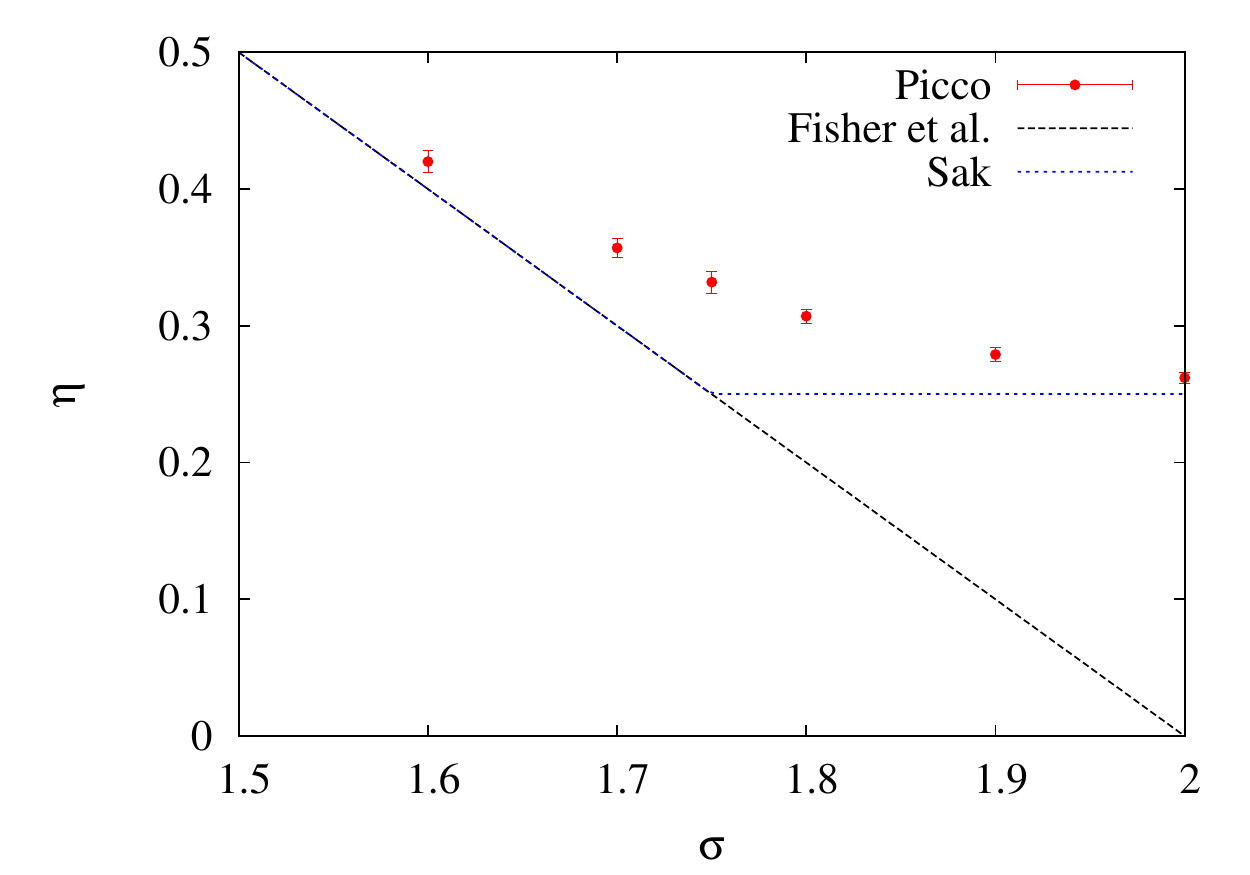}
\end{center}
\caption{\label{Fig:eta} Behaviour of $\eta(\sigma)$ for a LR model in $d=2$ as proposed in different works: in the work of Fisher et al.~\cite{fieldtheoryLR}	 $\eta=2-\sigma$ up to $\sigma=2$, while for Sak \cite{Sak} $\eta=\max(2-\sigma,\eta_{SR}=\frac{1}{4})$, and the data of Picco \cite{Picco} $\eta$ seem to interpolate smoothly between $2-\sigma$ and $\eta_{SR}=\frac{1}{4}$.}
\end{figure}

In Ref.~\cite{Sak} a different scenario was proposed. In fact, if the term $j_2k^2$ is not ignored in Eq. (\ref{Eq:LRLagrangian}) when $\sigma<2$,
as done in Ref.~\cite{fieldtheoryLR}, it can be seen that the non-trivial fixed point is characterized by $j_2^*=O(\epsilon^2)\neq0$.
Even if one starts with $j_2=0$, SR forces appear after the renormalization, determining the critical behaviour.
As a consequence, for $d < 4$ the boundary between the intermediate and the SR regime was found to shift from
$ \sigma_L= 2$ to $\sigma_L = 2-\eta_{SR}$.
In particular, for $\sigma<2-\eta_{SR}$, the introduction of $j_2\neq0$ does not change the critical exponents, that remain those of
Ref.~\cite{fieldtheoryLR}. When $\sigma>2-\eta_{SR}$, all the exponents become the SR ones, without discontinuity, and without loosing
the monotonicity in $\sigma$.
In fact in this regime the fixed point is characterized by $j_{\sigma}^*=0$, and the field theory is the usual SR one.

In support to this picture, in a field-theoretic approach, Honkonen and Nalimov \cite{Honkonen} proved, 
to all orders in perturbation theory, the stability of the SR fixed point for $\sigma> 2-\eta_{SR}$
and of the LR one for $\sigma< 2-\eta_{SR}$. 
Within this new scenario, the theory is also consistent with the exact results for the one-dimensional case.
In fact, for $d=1$, $\eta_{SR}=1$. In this way the lower critical exponent is $\sigma_L=1$, as expected.
However, the analysis of \cite{Sak} has also been the subject of
criticism: in Ref.~\cite{vanEnter} the results for $n\geq2$ are contested, 
in Ref.~\cite{Yamazaki} the absence of the kink at $\sigma=2-\eta_{SR}$ is hypothesized, in Ref.~\cite{Gusmao}
the picture of Ref.~\cite{fieldtheoryLR} is supported.
All this works on the subject are related to the importance of understanding how to treat systems in presence of different, 
competing fixed points.

There are also numerical studies. 
In Ref.~\cite{LR_SRnumeric} a Monte Carlo study of a LR model in $d=2$, using cluster algorithms, supports the scenario of Ref.~\cite{Sak} where 
$\eta=\max(2-\sigma,\eta_{SR}=\frac{1}{4})$, 
excluding definitively the picture of Ref.~\cite{fieldtheoryLR}.
In particular they affirm to find logarithmic corrections to scaling at $\sigma=1.75$, 
clear indication of a crossover between different critical points.
Very recently, in Ref.~\cite{Picco} the same study has been improved. 
In fact, the measurement of $\eta$ for a LR system in $d=2$ has been repeated,
close to the region where its behavior is changing, i.e. for $\sigma\simeq2 -\eta_{SR}$, obtaining more precise results. 
The author of Ref.~\cite{Picco} confirms that there is no
discontinuity but a clear deviation from the behavior predicted by Sak in Ref.~\cite{Sak} is measured.
In particular in the intermediate regime up to $\sigma \simeq 1.5$ the results are
in agreement with the prediction $\eta=\eta_{LR}=2-\sigma$. For $\sigma>2$, $\eta$ is in perfect agreement with the
value for a SR model. In the remaining part for $1.6 \leq\sigma\leq 2$ the results do not agree
with the prediction of the RG analysis \cite{fieldtheoryLR}\cite{Sak}. On the
contrary, $\eta$ seems to interpolate smoothly between these two behaviors. 
This behaviour is also supported by a recent RG calculation \cite{Picco2}.
Moreover logarithmic corrections are not found in this region.
The results in Ref.~\cite{LR_SRnumeric} are compatible with those in Ref.~\cite{Picco}, due to the larger error bars.
Concluding, the scenario at the lower critical exponent is far from being clear. 
The three proposed behaviours are summarized in Fig. \ref{Fig:eta}.

A somehow related problem is the identification of a \sDr relation that links the exponent $\sigma$
of the LR model in $d$ dimensions 
with the dimension $D$ of an effective SR model with the same critical behavior.
Comparing the field theory of a LR model in one dimension ($d=1$) with that of a SR model above its upper critical dimension ($D>D_U)$, 
the relation 
\begin{equation} \label{Eq:sigma}
\sigma=\frac{2}{D}
\end{equation}
is found.
The upper critical dimension for the FM SR model, $D_U=4$, 
thus corresponds to the upper critical exponent for the FM LR model $\sigma_U=1/2$. 
In the same way, the upper critical dimension for the SG SR model, $D_U=6$, 
 corresponds to the upper critical exponent for the SG LR model $\sigma_U=1/3$. 
Moreover $\sigma=0$ corresponds to $D=\infty$, as one can expect. However, this relation has a problem. 
In fact, the exponent for which there is no more a phase transition, 
$\sigma_L=1$, will correspond to a lower critical dimension in SR models $D_L=2$. 
But we know that the lower critical dimension for a SR model (with discrete degrees of freedom) is $D_L=1$. 
This problem can be overcome modifying slightly the matching relation \cite{eta-D}:
\begin{equation} \label{Eq:sigma2}
\sigma=\frac{2-\eta_{SR}(D)}{D}
\end{equation}
and with scaling arguments the following relations between the critical exponents for LR in $d=1$ and SR models can be found \cite{eta-D_2012}:
\begin{equation}
\begin{aligned}
 &\nu_{LR}(\sigma)=D\nu_{SR}(D); \quad 2-\eta_{LR}(\sigma)=\sigma=\frac{2-\eta_{SR}(D)}{D};\\
 &\gamma_{LR}(\sigma)=\gamma_{SR}(D);\quad \omega_{LR}(\sigma)=\frac{\omega_{SR}(D)}{D}.
\label{Eq:D-sigma_exponents}
\end{aligned}
\end{equation}

Each relation among the four in Eq. (\ref{Eq:D-sigma_exponents}) defines a \sDr correspondence. 
If SR and LR models are in the same universality class, then all the four correspondence in Eq. (\ref{Eq:D-sigma_exponents}) are equivalent.

Please note that the useful aspect of our definitions of the models is that all the \sDr
relations are valid both for the FM and the SG versions of the models. 

If one wants to test the exactness of the equivalence between a D-dimensional SR model and a one-dimensional LR model, 
one has to simulate a LR system at a value of $\sigma$ that corresponds to $D$ following for example Eq. (\ref{Eq:sigma2}), 
and verify if there is the correspondence between all the exponents as in Eq. (\ref{Eq:D-sigma_exponents}).
For the FM there is not a systematic study of the correspondence, 
while, only during the writing of this work, this problem has been analyzed for SG in Ref.~\cite{eta-D_2012} for $D=3$ and $D=4$.
For $D=4$ the matching between LR and SR models seems very good. For $D=3$ the data are evenly compatible, however errors are bigger 
and the answer is not definite. 
Anyhow it is reasonable that the correspondence between SR and LR models becomes weaker approaching the lower critical dimension. 
Indeed at the upper
critical dimension the field theory is exactly the same, while at the lower 
critical dimension for the FM model we know that SR and LR models have even qualitatively different behaviours.
In fact the SR model has a $T=0$ transition, while the LR model has a KT transition. 

Another LR model, widely used is the Dyson hierarchical model (HM) \cite{Dyson} (see \cite{Meurice} for a review).
It is a particular one-dimensional LR model, in which the Hamiltonian of $2^n$ spins can be constructed iteratively in the following way:
\begin{equation}
H_n(s_1,...,s_{2^n}) = H_{n-1}(s_1,...,s_{2^{n-1}}) +
+ H_{n-1}(s_{2^{n-1}+1},...,s_{2^n}) + c^n \sum_{i<j=1}^{2^n} J_{ij}\,s_i\,s_j\;.
\label{eq:Hindip}
\end{equation}
The intensity of the interactions decreases with the level $n$ by a factor $c=2^{-(\sigma+1)}$.
One expects the model to behave like the usual LR one, with the same exponent $\sigma$, because the decaying at large scales of the 
coupling intensity is the same.
Indeed the model undergoes a standard second order phase transition if $2^{-1}>c>2^{-2}$ \cite{Dyson} (i.e. $0<\sigma<1$). For $2^{-1}>c>c_U=2^{-3/2}$
(i.e. $0<\sigma<\sigma_U=1/2$), 
the Gaussian solution of the field theory associated to this model is the stable one 
and the critical exponents are the mean-field ones as for usual LR systems.
Again, for $2^{-3/2}>c>c_L=2^{-2}$ (i.e. $1/2<\sigma<\sigma_L=1$) the exponents differ from the classical ones, 
but nobody has checked if and how much they differ from the LR ones.
The first order term in the $\epsilon$-expansion of the two models is the same, while the second order one differs slightly,
the coefficients being 4.445 for the HM and 4.368 for the LR model \cite{nu_hier}.

One crucial difference between the two models is that for the borderline case $\sigma_L=1$ there is no KT phase-transition for the HM. 
Indeed in the HM all the interactions are weaker than in the usual LR model. 
For this reason if the HM has a transition, it implies that the LR model has a transition too, but the vice-versa is not necessarily true. 
Nonetheless, there is a KT phase transition also in the HM for $\sigma=1$ if interactions at level $n$, $J_n=2^{-2 n}$, are made slightly stronger, i.e.\ $J_n=2^{-2 n}\log(n)$ \cite{LCD-log-Dyson}.

\section{New results on the connection between LR and SR models}

\subsection{Monte Carlo algorithm and data analysis}

To the best of our knowledge there exist no estimates of the critical exponents in $d=1$ for values of $\sigma$ corresponding to $D=2$ and $D=3$ following Eq.~(\ref{Eq:sigma2}) neither for the power law LR models nor for the HM. 
For this reason we have performed Monte Carlo simulations at these values of $\sigma$.
Indeed Eq. (\ref{Eq:sigma2}) was introduced recently studying SG models, and was never applied to FM, 
for which the relation (\ref{Eq:sigma}) was often used. 
Moreover we want to see how similar is the HM with respect to the power law LR model.

We have simulated the $d=1$ LR model using the cluster algorithm proposed in Ref.~\cite{cluster_LR}, where the use of the cumulative probability distribution for adding a new spin to the cluster to be flipped allows to keep running times $O(N)$ even if the model is fully connected\footnote{Standard cluster algorithms usually require $O(N^2)$ operations for fully connected models.}.

In cluster algorithms, a first spin $\sigma_i$ is randomly chosen and the neighbors $\sigma_j$ having the same sign are inserted in the cluster with a probability 
\begin{equation}
p_j=1-e^{-2\beta J_{ij}}.  
\label{Eq:clusterprobability}
\end{equation}
The probability that the first neighbor to be included in the cluster is the $j$-th from the reference spin is
\begin{equation}
P(j)=p_j(1-p_{j-1})...(1-p_{1})\;.
\label{Eq:clusterP} 
\end{equation}
Thus, by defining the cumulative bond probability
\begin{equation}
C(j)=\sum_{n=1}^jP(n)\;,
\label{Eq:cumulative} 
\end{equation}
and extracting a random number $r$ uniformly in $[0,1]$, if $C(j-1)<r\leq C(j)$ 
then the first spin included in the cluster is the $j$-th. The condition on the spin being parallel 
to those in the cluster is checked after the
selection. If the selected spin is antiparallel to those in the cluster, it is not added.

After the first neighbor has been chosen, we want to include in the cluster spins at distance $k>j$. 
Eq.~(\ref{Eq:clusterP}) is generalized to
\begin{equation}
P_j(k)=p_k(1-p_{k-1})...(1-p_{j+1})\;,
\end{equation}
and it leads to a cumulative bond probability
\begin{equation}
C_j(k)=\sum_{n=j+1}^kP_j(n)=1-\exp\(({\sum_{n=j+1}^k-2\beta J_n}\))\;,
\label{Eq:cumulative2} 
\end{equation}
where Eq.~(\ref{Eq:clusterprobability}) has been used to obtain the last expression 
($J_n$ is the coupling between spins at distance $n$).
A new random number is extracted and a new spin is selected. Spins are added in this way until the maximum distance $N/2$ is reached.
Then we try to add neighbors starting from all the other spins already inserted in the cluster in the same way.
Naturally in this procedure we have to take into account that there are more than one spin at distance $k$ 
(especially in dimensions higher than 1),
and we must ensure that every spin is counted with the right probability.

Given $C_j(k)$ we construct a look-up table to calculate the distance $k$ associated to the random number extracted.
In this way the cumulative probability is calculated only once at the beginning and it is the same for all the spins, 
since the system is homogeneous.
Moreover only $C(j)$ has to be computed, since $C_j(k)$ can be derived from it as
$$
C_j(k)=\frac{C(k)-C(j)}{1-C(j)}\;.
$$
Once the random number is extracted, we search in the look-up table to determine $k$. This operation has a cost $O(\log(N))$.
The main advantage of this method is that it is exact, at variance to the one of Ref.~\cite{cluster_LR}.

We have used periodic boundary conditions, such that two spins $i$ and $j$ interact 
with a single coupling that depends on the minimum distance between them: 
$r_{ij}=\min(|i-j|, L-|i-j| )$, where $L$ is the size of the system.

We have performed Monte Carlo simulations of at least $10^6$ MCS.
We have checked for the equilibration dividing the measurements in bins with a geometrically growing size, and we have assumed that the system has reached the equilibrium when the average of the magnetization in at least the last two bins is the same within the error (that is at least $3/4$ of the simulation is sampling the same average magnetization).
We have found that the equilibration time is $\tau \simeq 10^5$ cluster MC steps for the largest sizes.
Willing to compute the susceptibility and the Binder parameter, we need the second and fourth moments of the magnetization.
We have obtained two different estimates for these quantities. The first estimate is the usual one:
$$
m^2=\<\Big(\frac{1}{N}\sum_i \sigma_i\Big)^2\>\;,\qquad
m^4=\<\Big(\frac{1}{N}\sum_i \sigma_i\Big)^4\>\;.
$$
The second method uses the \textit{improved estimators} that can be defined when cluster algorithms are used \cite{improved_estimators}:
\begin{equation}
m^2=\frac{1}{N}\<|C|\>\;, \qquad
m^4=\frac{3}{N^2}\<|C||C'|\>-\frac{2}{N^3}\<|C|^3\>\;.
\label{Eq:improved}
\end{equation}
where $|C|$ and $|C'|$ are the sizes of flipped clusters.
Operatively, we compute $\<|C||C'|\>$ in the following way: we choose randomly a spin and, starting from it, we construct a cluster. 
We call $|C|$ the number of spins of this first cluster. Then we choose a second spin randomly. 
If it is in the cluster we already built, we put $|C'|=|C|$.
If it is not in that cluster, we construct a new cluster starting from it, and we call the new cluster size $|C'|$.
$C$ and $C'$ are always disjoint (i.e., non overlapping).
Please note that we can not compute the average $\<|C||C'|\>$ simply as $\<|C|^2\>$  because in this way we would not take into account the condition $|C|+|C'|\leq N$.

While the improved estimator for $m^2$ in Eq.~(\ref{Eq:improved}) has been already introduced in Ref.~\cite{improved_estimators}, we believe the one for $m^4$ is new.
We have computed the susceptibility and the Binder parameter and their errors with the jackknife method separately for the two methods.
At the end we have taken the weighted average between the two values. 
In this way we are conscious that we are underestimating a little the error 
because the two measures are correlated but we assume them to be uncorrelated when we perform the weighted average.

We have used this method to simulate a one-dimensional LR model with values of $\sigma$ corresponding to $D=2$ and $D=3$.
In $D=2$ we know exactly the exponent $\eta=\frac{1}{4}$ and it corresponds to $\sigma=\frac{2-1/4}{2}=0.875$. 
In $D=3$, $\eta=0.0364(5)$ as found in Ref.~\cite{Pellissetto} and it corresponds to $\sigma=\frac{2-0.0364}{3}=0.65453$. 
We have computed the critical exponents $\nu$ and $\omega$ using a Finite Size Scaling (FSS) analysis.

A great advantage of LR models is that the $\eta$ exponent is not renormalized in the non-mean-field region as explained before; thus we know its analytical expression: $\eta=2-\sigma$.
For this reason we can compute from the susceptibility $\chi=\frac{N}{T}\<m^2\>$ the scale-invariant quantity $\chi_L/L^{\sigma}$.
Another quantity that we look at is the dimensionless Binder parameter: 
$B=\frac{1}{2}\Big[3-\frac{\langle m^4 \rangle}{\langle m^2 \rangle^2}\Big]$.
Both observables should cross at $T_c$ for large sizes.

\begin{figure}[t]
\begin{center}	
\includegraphics[width=0.7\columnwidth]{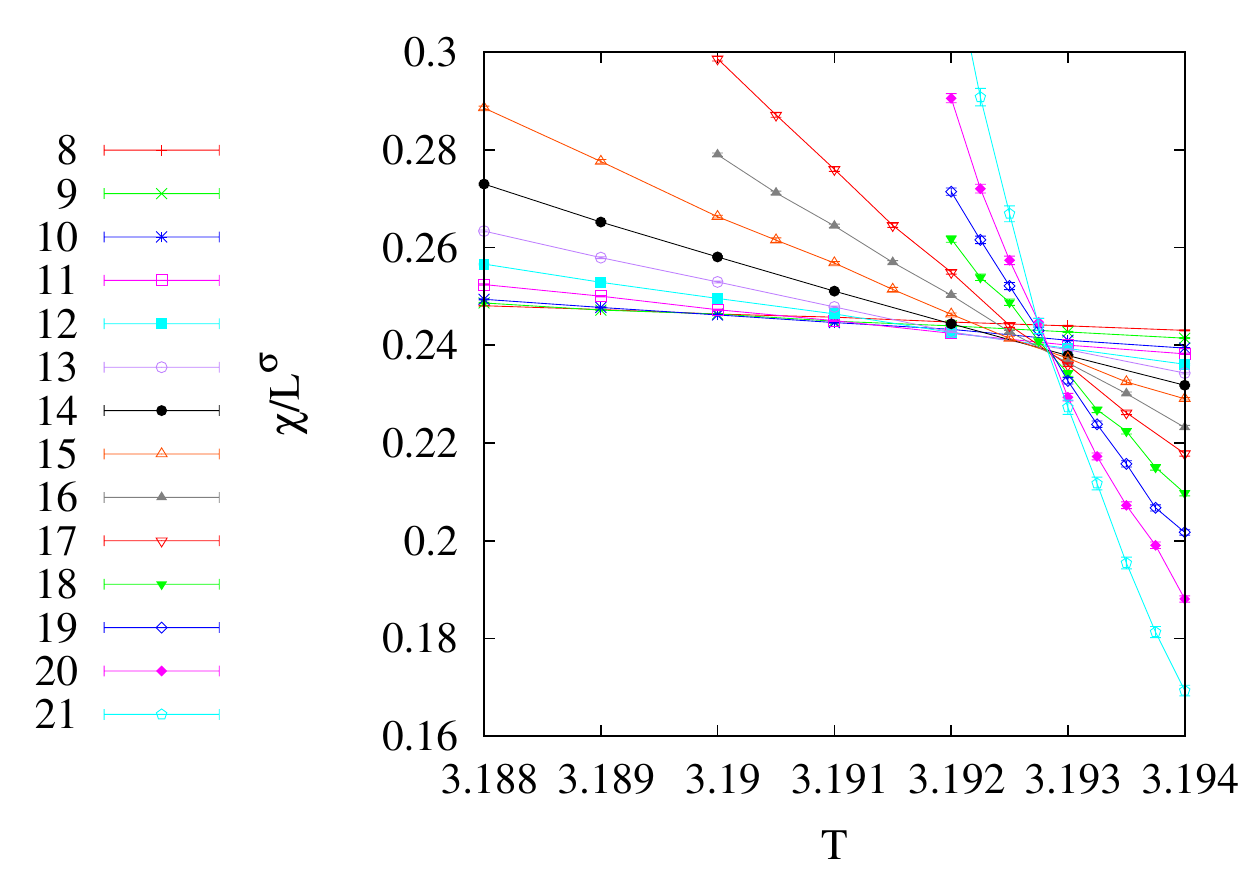}
\includegraphics[width=0.7\columnwidth]{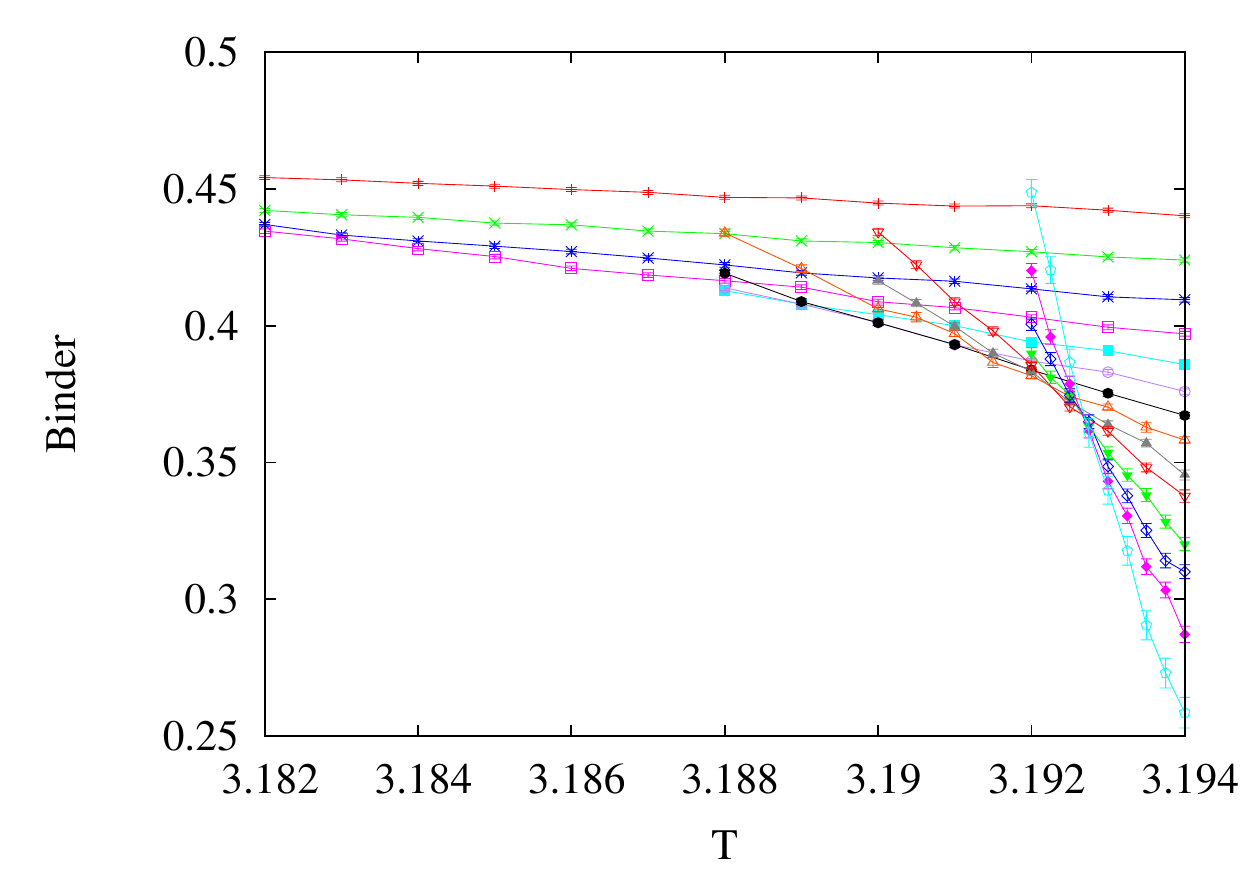}
\end{center}
\caption{\label{Fig:binder_chi} Scale invariant observable $\chi_L/L^{\sigma}$ (left) and Binder cumulant (right), computed at different $n=\log_2L$, as a function of the temperature $T$, at $\sigma=0.654533$.
The curves at different sizes should cross at a temperature that approaches $T_c$ when $L$ grows. The Binder cumulant shows stronger corrections to scaling with respect to $\chi_L/L^{\sigma}$.}
\end{figure}

\begin{figure}[t]
\begin{center}	
\subfloat{\includegraphics[width=0.5\columnwidth]{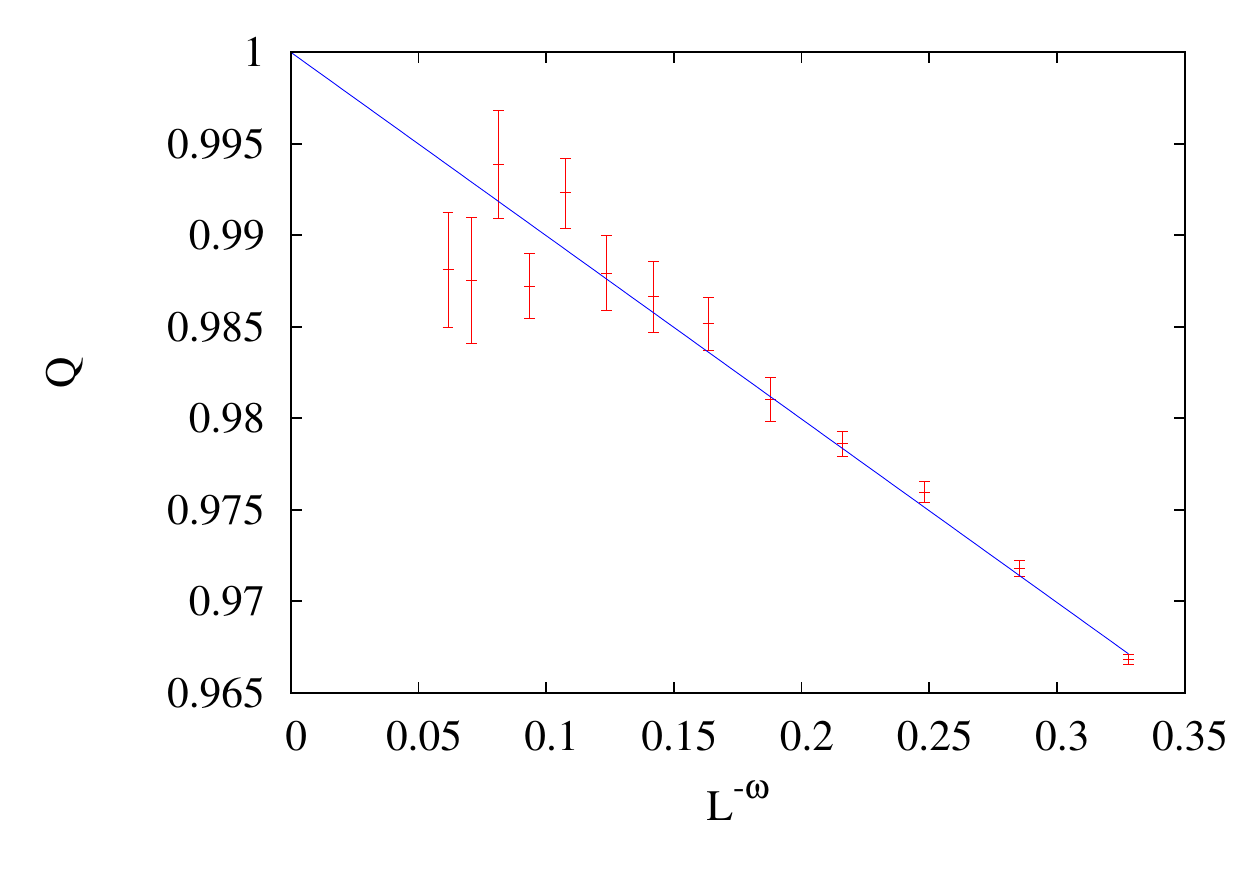}}
\subfloat{\includegraphics[width=0.5\columnwidth]{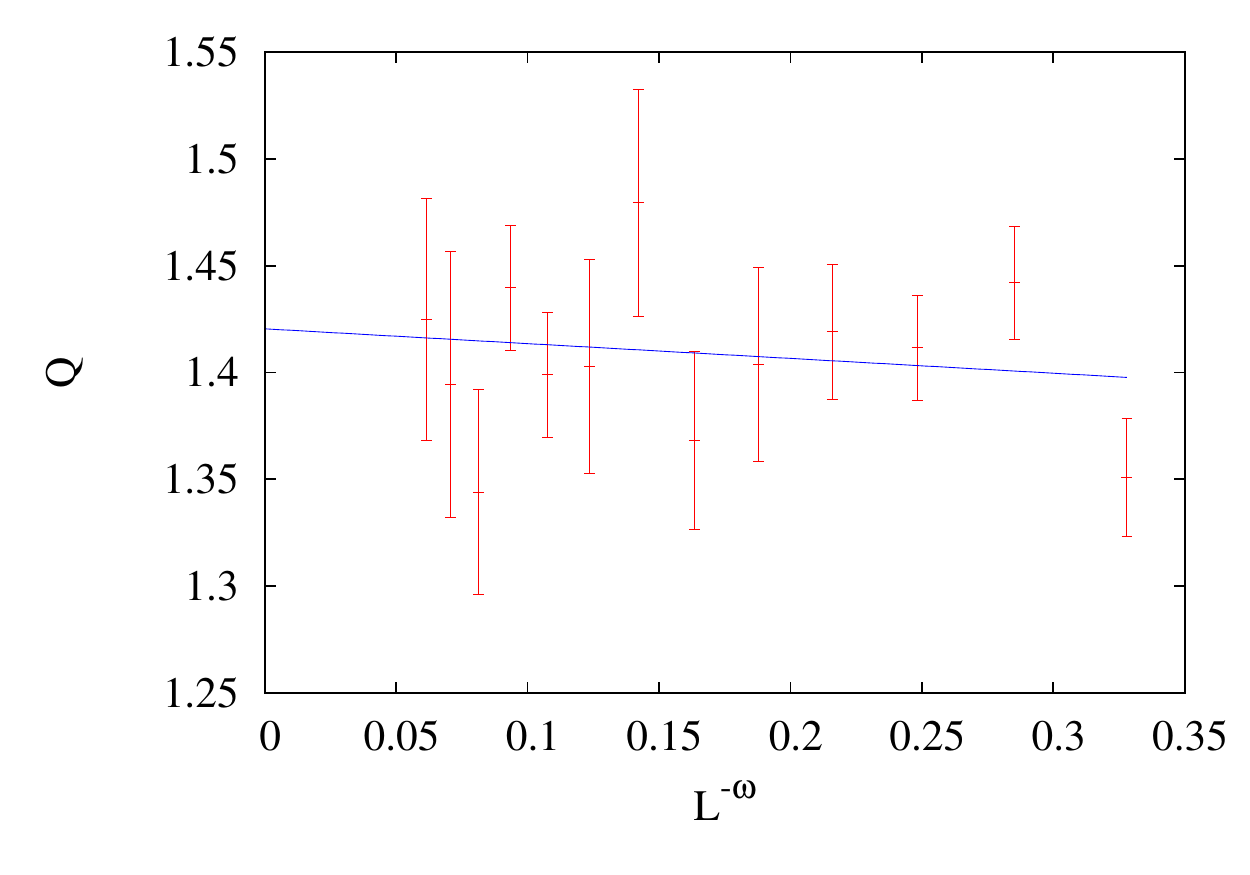}}
\end{center}
\caption{\label{Fig:omega} Left: quotient of the Binder parameter for $L$ and $2L$ at $T^*_L$, 
computed at the crossing temperature of $\chi_L/L^{\sigma}$.
The straight line is the best fit using Eq. (\ref{Eq:omega}) with $\omega$ left as a free parameter.
Right: quotient of the derivative of the Binder parameter at $T^*_L$. 
The straight line is the best fit as a function of $L^{-\omega}$ using Eq. (\ref{Eq:nu}), 
with $\omega$ determined from the previous fit and the intercept $2^{1/\nu}$ left as a free parameter.}
\end{figure}

In Fig. \ref{Fig:binder_chi} the two observables $\chi_L/L^{\sigma}$ and $B_L$ are plotted as a function of the temperature, 
around the critical 
temperature $T_c$, for different sizes $L=2^n$ of the systems. 
We have extracted the temperatures $T^*_L$ of the crossing of $\chi_L/L^{\sigma}$ for sizes $L=2^n$ and $L'=2L=2^{n+1}$.
They should approach the critical point following: 
\begin{equation}
T^*_L=T^*_{\infty}(1+aL^{-\omega-\frac{1}{\nu}}).
\label{Eq:T_c}
\end{equation}
We have computed the values of the Binder parameter $B(L,T^*_L)$ and the quotient $Q=\frac{B(2L,T^*_L)}{B(L,T^*_L)}$ at the previously extracted temperatures $T_L^*$.
The latter behaves as:
\begin{equation}Q=\frac{B(2L,T^*_L)}{B(L,T^*_L)}=1+bL^{-\omega}.
 \label{Eq:omega}
\end{equation}
Thus we have performed a fit with $\omega$ left as a free parameter.
The results are shown in the left side of Fig. \ref{Fig:omega}.
Once we have determined $\omega$, we extract the derivative of the Binder parameter at $T_L^*$, $B'(L,T^*_L)$, as the angular coefficient of the straight line passing through the data. We compute the quotient $Q=\frac{B'(2L,T^*_L)}{B'(L,T^*_L)}$ that follows:
\begin{equation}
 \frac{B'(2L,T^*_L)}{B'(L,T^*_L)}=2^{1/\nu}+cL^{-\omega}.
\label{Eq:nu}
\end{equation}
Thus, using the value of $\omega$ previously determined and performing a linear fit as a function of $L^{-\omega}$, 
we extract the
value of $\nu$ from the intercept. The results are shown in the right panel of Fig. \ref{Fig:omega}.
At this point, fitting with a line the values of $T^*_L$ as a function of $L^{-\omega-1/\nu}$, 
with the previously determined $\omega$ and $\nu$, 
we can extract $T_c$ as the intercept, using Eq.~(\ref{Eq:T_c}).

\subsection{Results for the 1d LR models}

The results in $d=1$ for $\sigma=0.875$ (corresponding to $D=2$) and $\sigma=0.654533$ (corresponding to $D=3$) are the following:
$$\frac{1}{\nu_{LR}(0.875)}=0.4124(13), \quad T_c(0.875)=2.10589(1)$$
$$\frac{1}{\nu_{LR}(0.65453)}=0.506(14) ,\quad \omega_{LR}(0.65453)=0.201(11) ,\quad T_c(0.65453)=3.19289(2) .$$
For $\sigma=0.875$ it is quite impossible to determine $\omega$ because we see very little dependence of $T^*_L$ with $L$ and the quotient of the Binder parameter is nearly independent of the size.

For the HM, we performed the same analysis. The only difference is that we computed the Binder parameter 
and the susceptibility exactly using the exact recursion relation for the probability of the magnetization at level $n$	:
\begin{equation*}
p_n(m) \propto e^{\beta c^n m^2}\!\sum_{m_L,m_R} p_{n-1}(m_L)\,p_{n-1}(m_R)\,\delta_{m_L+m_R,m}
\label{eq:Pm}
\end{equation*}
where $m_L$ and $m_R$ are the magnetizations of the half systems.
The results are:

$$\frac{1}{\nu_{HM}(0.875)}=0.3841(9),\quad \omega_{HM}(0.875)= 0.462(3)$$
$$\frac{1}{\nu_{HM}(0.65453)}=0.5186(72),\quad \omega_{HM}(0.65453)=0.212(5).$$
Naturally for the HM, many other methods can be used to obtain more precise results. 
However, for the exponent $\omega$ only an estimate is available 
\cite{omega-HM}
$$\omega_{HM}(2/3)=0.2185787,$$ consistent with our results.

The values for the $\nu$ and $\omega$ exponents of LR and HM for $\sigma=0.65453$ are in perfect agreement, while those for $\sigma=0.875$ differ.
Moreover if we compare them with the SR values \cite{Pellissetto}
$$\omega_{SR}(D=3)=0.84(4) , \quad \omega_{SR}(D=2)=2$$ and remembering the supposed relation between them, 
$\omega_{SR}(D)=D\omega_{LR}(\sigma)$, it seems that LR models have bigger finite size effects (smaller $\omega$) than SR models.
Thus, looking at the $\omega$ exponent, Eq. (\ref{Eq:D-sigma_exponents}) is not satisfied, especially for $D=2$.

If we compare the values for the $\nu$ exponent with the SR ones, 
$\nu_{SR}(2)=1$ and $\nu_{SR}(3)=0.6301(4)$ \cite{Pellissetto}, 
we see that Eq.~(\ref{Eq:D-sigma_exponents}) is a good approximation even for $D=3$ (that is near enough to the upper critical dimension)
 [$\nu_{LR}=1.976(55) \simeq 3 \nu_{SR} = 1.8903(12)$],
but it is no more good for $D=2$ [$\nu_{LR}=2.425(8) \neq 2 \nu_{SR} = 2$].

\subsection{Generalization of the \sDr relations in more than one dimension}

How can we generalize the \sDr relations if the LR model is defined in more than one dimension? 
Let us first remark the notation: $d$ is the real dimension of the LR model while $D$ is the dimension of an equivalent SR model.
If we use the same arguments of the one-dimensional case in Ref.~\cite{eta-D_2012} for the scaling form of the free energy, 
the relation
\begin{equation} \label{Eq:sigmaDd}
\frac{\sigma}{d}=\frac{2-\eta_{SR}(D)}{D}
\end{equation}
is obtained.

However Eq. (\ref{Eq:sigmaDd}) can be also obtained from another way.
In fact one can think that an approximate super-universality exists. 
The conjecture is that the exponent $\gamma_{LR}(d,\sigma)$ and other quantities are approximately functions only of 
$\hat{\sigma}=\sigma/d$. 
This conjecture is exact in all the mean-field region. In fact, $\gamma_{LR}=1$ in the region $0<\hat{\sigma}<\frac{1}{2}$, 
independently on $d$.
The SR model is recovered when $\sigma=\sigma_L(d)=2-\eta_{SR}(D=d)$ \cite{LR_SRnumeric,Sak}. 
If now we use this information, we obtain that $\gamma_{SR}(D)=\gamma_{LR}(\hat{\sigma}=\frac{2-\eta_{SR}}{D})=\gamma_{LR}(\frac{\sigma}{d})$. 
Thus the new relation between a LR model in $d$ dimensions and an effective SR model in $D$ dimensions is Eq. (\ref{Eq:sigmaDd}).
In this way we have connected two problems: the determination of the \sDr relation and the threshold $\sigma_L$ 
where the SR behaviour is recovered.
These problems are often viewed as disconnected, however we think that they are closely related.

Please note that the value of $\sigma_L$ is not universal:
$\sigma_L(1)=1, \sigma_L(2)=\frac{7}{4}$ \cite{Onsager}, $\sigma_L(3)=2-0.0364=1.9636$ \cite{Pellissetto}, $\sigma_L(4)=2$.
In the same way, $\hat{\sigma}=\hat{\sigma}_L$ is not universal: 
$\hat{\sigma}_L(1)=1$ , $\hat{\sigma}_L(2)=0.875$ , $\hat{\sigma}_L(3)=0.65453$, $\hat{\sigma}_L(4)=0.5$.
For the exponent of the correlation length, using the scaling relation $\nu=\gamma/(2-\eta)$, 
the known value of $\eta=2-\sigma$ in the LR region and Eq. (\ref{Eq:sigmaDd}), 
one obtains:
\begin{equation} \label{Eq:nu-relation}
\nu_{SR}(D)=\frac{\gamma_{SR}(D)}{2-\eta_{SR}(D)}=\frac{d}{D\sigma}\gamma_{LR}\left(\frac{2-\eta_{SR}(D)}{D}\right)=\frac{d}{D}\nu_{LR}\left(\frac{2-\eta_{SR}(D)}{D}\right)
\end{equation}

In analogy with Eq. (\ref{Eq:D-sigma_exponents}), one can thus suppose that there exists a value of $\sigma$ 
that satisfies all the following relations for the critical exponents:

\begin{equation}
\begin{aligned}
 &d\,\nu_{LR}(\hat\sigma)=D\,\nu_{SR}(D); \quad \frac{2-\eta_{LR}(\hat\sigma)}{d}=\frac{2-\eta_{SR}(D)}{D};\\
 &\gamma_{LR}(\hat\sigma)=\gamma_{SR}(D);\quad \frac{\omega_{LR}(\hat\sigma)}{d}=\frac{\omega_{SR}(D)}{D}.
\label{Eq:Dd-sigma_exponents}
\end{aligned}
\end{equation}
Please note that the two dimensions $d$ and $D$ enter only through their ratio.

\subsection{Simulations in $d=2$}

Unfortunately there were not previous estimates for the $\nu$ exponent for the LR model in more than one dimension, 
only during the completion of this work
in Ref.~\cite{Picco} the value $\nu_{LR}=0.96(2)$ for
$\sigma = 1.6$ in $d=2$ was reported, extracted from a Monte Carlo simulation.
For this reason we have also done simulations in $d=2$ to extract the exponents at values of $\sigma=1.20, 1.60$ 
and in particular at $\sigma=1.30906$ 
that corresponds to $D=3$ and $\sigma=1.75$ where the SR behaviour in $D=2$ should be recovered.
For $\sigma=1.2$ and $\sigma=1.30906$, the simulations have been performed with the same cluster algorithm 
and the same analysis method 
as for $d=1$.
The  obtained values for the $\nu_{LR}(\sigma)$ and $\omega_{LR}(\sigma)$ exponents and for the critical temperatures are:
$$\frac{1}{\nu_{LR}(1.2)}=1.024(34),\quad \omega_{LR}(1.2)=0.480(25) , \quad T_c(1.2)=6.83427(1),$$ 
$$\frac{1}{\nu_{LR}(1.30906)}= 1.014(33),\quad \omega_{LR}(1.30906)=0.32(15), \quad T_c(1.30906)=6.32546(4) .$$  
The value of $\nu_{LR}$ at $\sigma=1.30906$ is compatible with the one for the $D=3$ SR model $\nu_{SR}=0.6301(4)$ \cite{Pellissetto} 
following Eq. (\ref{Eq:Dd-sigma_exponents}): 
$2 \nu_{LR} = 1.97(6) \simeq 3 \nu_{SR} = 1.8903(12)$.
The value of $\omega_{LR}$ at $\sigma=1.30906$ is very difficult to extrapolate because there are unusual non-monotonic finite size effects.

For $\sigma=1.6$ and $\sigma=1.75$ the finite size effects look extremely strong. 
Indeed the size-dependent critical temperatures where the Binder cumulants cross drifts a lot by varying the system size, and it is not possible to extract the critical Binder value $B(\infty,T_c)$.
For this reason we move to use a slightly different model, where the sum over all images is made as in  \cite{LR_SRnumeric}, thus leading to new couplings
$$J_{ij}=\sum_{x=-\infty}^{\infty}\sum_{y=-\infty}^{\infty}\(((x_i-x_j+Lx)^2+(y_i-y_j+Ly)^2\))^{-(d+\sigma)/2}\;.$$
In the thermodynamic limit the two models (with and without images) are equivalent.
In Ref.~\cite{LR_SRnumeric} the authors were able to compute the contributions from all the images exactly, because they used slightly different couplings defined as
\begin{equation}
J_{ij}=\int_{|i-j|-\frac{1}{2}}^{|i-j|+\frac{1}{2}} x^{-(d+\sigma)} dx\;.
\label{Eq:coupling_integral} 
\end{equation}
Since we use the original definition of the couplings, in principle it would not be possible to include all the images exactly.
To overcome this problem, we estimate the error that we commit on the largest coupling 
(formally the one between two spins at distance 0) by including only the first $(2a)^2$ images (that is images within a distance $a$):
$$
\int_{\substack{|x|>a\\|y|>a}}\text{d}x\,\text{d}y\(((Lx)^2+(Ly)^2\))^{-\frac{2+\sigma}{2}}<2\pi L^{-(2+\sigma)}\int_a^{\infty}\text{d}r\, r^{-(1+\sigma)}=\frac{2\pi L^{-(2+\sigma)}}{\sigma a^\sigma},$$
and we choose $a$ such as to make this error smaller than $10^{-9}$.
At this point we compute the new couplings between any pair of spins as the sum of the couplings between the $(2a)^2$ images.
Due to the large values of $\sigma$, the number of images considered is always small. If $a$ results to be smaller than 10, 
we choose $a=10$.
Adding the images, the observables show a reduced dependence on the system size, and the data analysis is cleaner.

We have not used the scale-invariant quantity $\chi_L/L^{2-\eta}$ because there is not agreement on the values of $\eta$ in this region. For this reason we have performed the following analysis.
We have looked at the temperatures $T^*_L$ at which the Binder cumulants for sizes $L$ and $2L$ cross. 
These crossings scale according to Eq.~(\ref{Eq:T_c}). Then we have fitted the values of $B(L, T^*_L)$ 
with a power law function of the type
\begin{equation}
 B(L,T^*_L)\simeq B(\infty)+aL^{-\omega},
\label{Eq:scaling_omega}
\end{equation}
determining $\omega$. For $\sigma=1.75$, assuming that the Binder parameter at the critical point should recover the SR value, we have used the value of the Binder parameter $B_{\infty}(T_c) = 0.91588...$ \cite{Binder2D} in the fit to reduce the uncertainty in the determination of $\omega$.
At this point we have computed the quotient of the derivative of the Binder parameter at $T^*_L$ and extracted the exponent $\nu$. Knowing $\nu$ and $\omega$, we have estimated $T_c$.

The obtained results are the following:
$$\frac{1}{\nu_{LR}(1.6)}=0.996(33),\quad \omega_{LR}(1.6)=0.130(45), \quad T_c(1.6)=5.29321(4),$$ 
$$\frac{1}{\nu_{LR}(1.75)}=0.98(10),\quad \omega_{LR}(1.75)=0.213(8), \quad T_c(1.75)= 4.89455(17).$$
We do not observe any logarithmic corrections, as already noticed in Ref.~\cite{LR_SRnumeric}.
The value of $\omega_{LR}$ at $\sigma=1.75$ is much smaller than that for a SR model in $D=2$, 
an explanation will be given in the following.
The value for $\nu_{LR}(1.75)$ is compatible with the SR one, $\nu_{SR}(D=2)=1$.

\subsection{At the lower critical $\sigma_L$}

\begin{figure}[ht]
\begin{center}	
\includegraphics[width=\columnwidth]{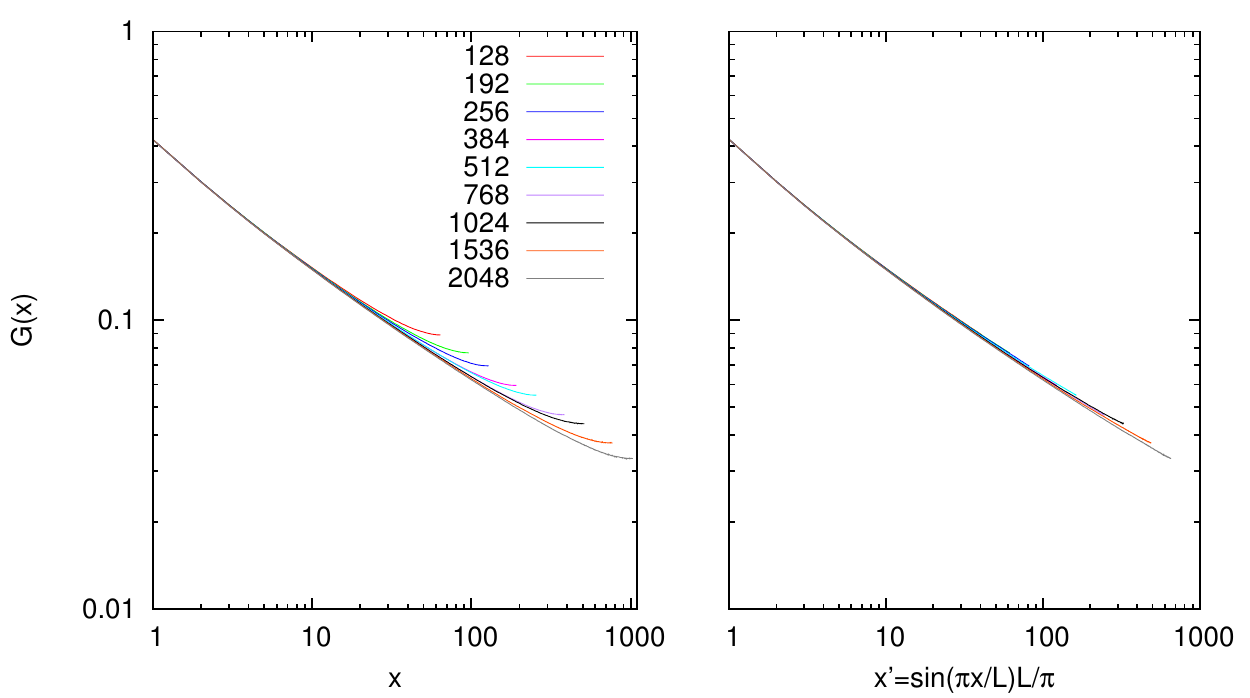}
\end{center}
\caption{The spin-spin correlation function for different sizes at $\sigma=1.75$ and $d=2$.
Left panel shows raw data, while in the right panel	boundary effects have been drastically reduced by plotting parametrically versus the variable $x'(x)\equiv\sin(\pi x/L)L/\pi$.}
\label{Fig:corr}
\end{figure}

In addition to the verification of the \sDr relations, 
we have concentrated our attention to the problems arising when approaching 
the value of $\sigma$ where the SR behaviour should be recovered. 
In particular we want to verify whether the scenario of Ref.~\cite{Sak} holds, with  
$\eta=\max(2-\sigma,\eta_{SR}=\frac{1}{4})$, or if, for $1.6 \leq\sigma\leq 2$, 
the $\eta$ exponent interpolates smoothly between $2-\sigma$ and $\eta_{SR}$ 
as stated in Ref.~\cite{Picco}. 
We notice that if the second scenario holds, 
the superuniversality conjecture can not be verified in the region near to $\sigma_L(d)$
where the $\eta$ exponent interpolates smoothly between the two behaviours.
Superuniversality is compatible only with the first scenario.

We have tried to measure $\eta$ in $d=2$ at the lower critical value $\sigma=\sigma_L=1.75$. 
We have performed MC simulations with a single image, because the use of images has the disadvantage that couplings slightly depend on the system size and consequently the small distance behavior of the correlation function does depend on the system size, making the study of finite size effects more complicated.
We have looked at the two-points correlation function at the critical point
that decays at large distances as $G(x)=\<\sigma(0)\sigma(x)\>=|x|^{-(d-2+\eta)}=|x|^{-\eta}$.
As it is customary, we have measured the correlation function along the principal axis:
$$
G(x)=\frac{1}{2N^2}\sum_{i,j}\Big(\<\sigma_{i,j}\sigma_{i+x,j}\>+\<\sigma_{i,j}\sigma_{i,j+x}\>\Big)\;.
$$
In the left panel of Fig.~\ref{Fig:corr} we plot the spin-spin correlation function for different sizes.
We notice that the effects due to the periodic boundary conditions, that actually imply the condition $G'(L/2)=0$, are rather severe and make hard to interpolate the data.
However the use of the variable $x'(x)\equiv\sin(\pi x/L)L/\pi$, that is actually an identity $x'(x)=x$ for $x\ll L$, is able to reduce drastically such boundary effects (see the right panel in Fig.~\ref{Fig:corr}). In the rest of the analysis we will use the rescaled variable $x'$, which is equal to $x$ in the thermodynamic limit, but allows for a better fitting of data at finite values of $L$.

The correlation function $G(x)$ at $\sigma=1.75$ can not be interpolated by a single power law: as shown in the right panel in Fig.~\ref{Fig:corr}, since $G(x)$ seems to decay faster at small distances and slower at large distances.
The same feature is not present at smaller $\sigma$, near to the upper critical value $\sigma_U=1$, nor in the SR model in $D=2$.
What we are observing is not a finite-size effect because it persists at large sizes.

\begin{figure}[ht]
\begin{center}	
\includegraphics[width=0.49\columnwidth]{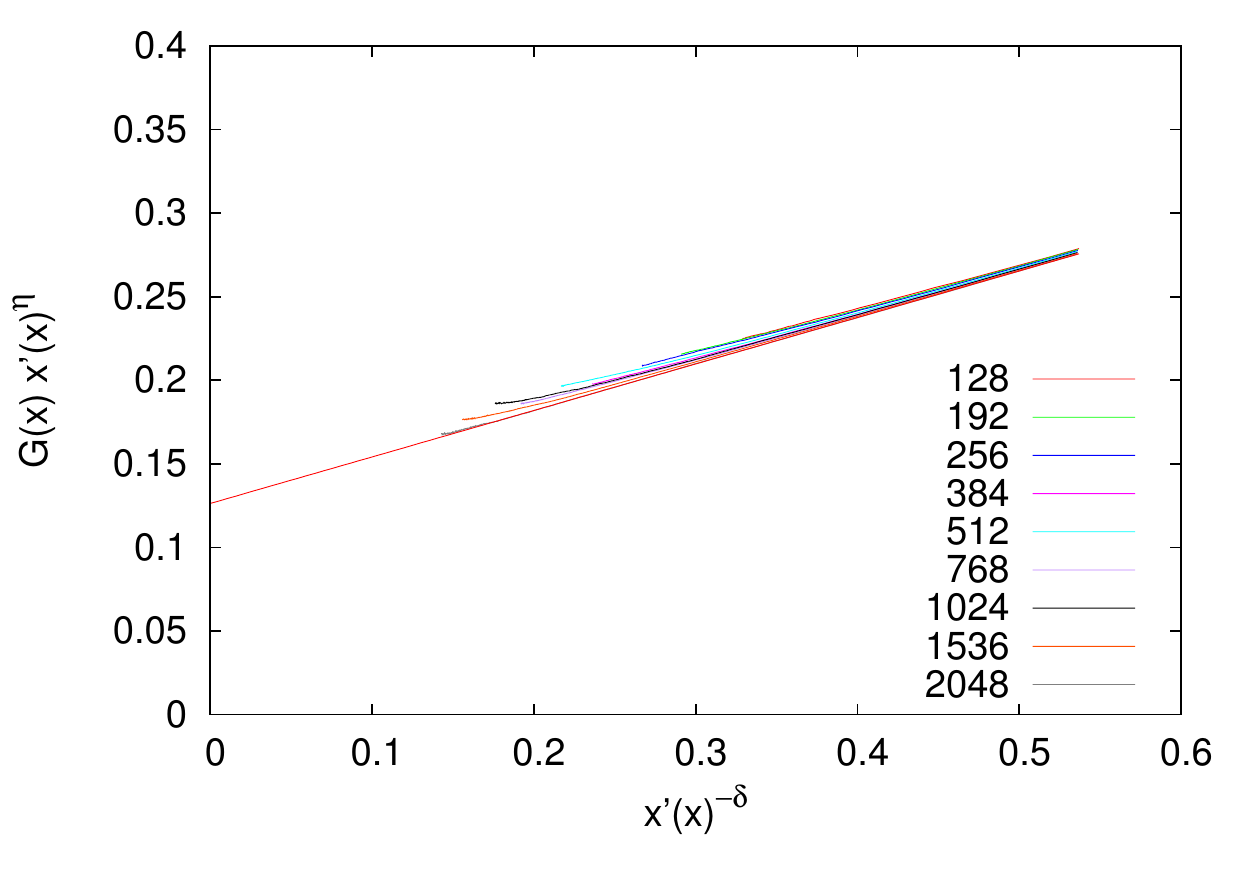}
\includegraphics[width=0.49\columnwidth]{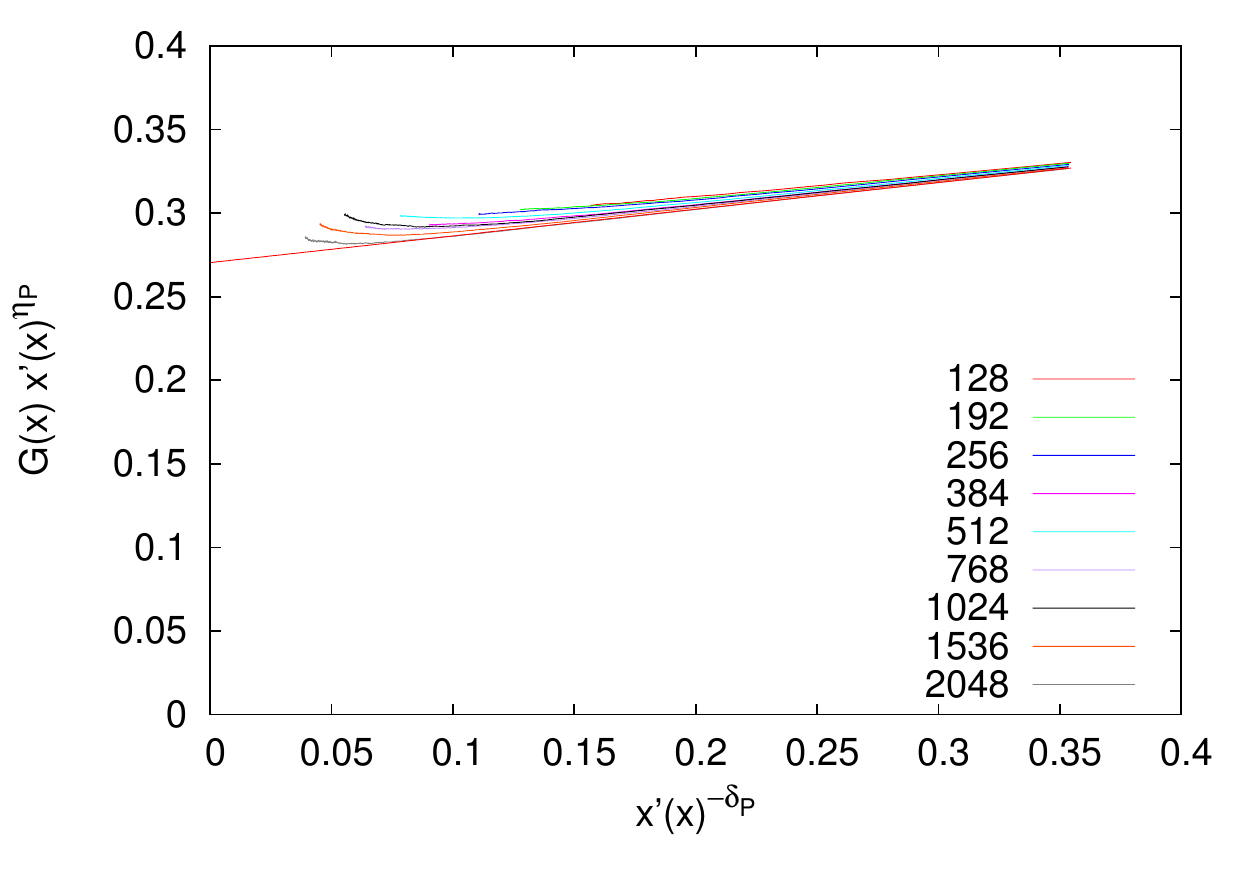}
\end{center}
\caption{Data for the spin-spin correlation function $G(x)$ measured at $\sigma=1.75$ ad $d=2$, rescaled by the asymptotic power law $x'(x)^\eta$, in order to highlight the corrections to the asymptotic decay.
Left: $\eta=0.25$ and $\delta=0.3$. Right: $\eta_P=0.332$ and $\delta_P=0.5$.}
\label{Fig:corr_s175_scaled}
\end{figure}

Since the $G(x)$ shows a small, but clear, upward curvature in a log-log scale we have interpolated the data through the following function
\begin{equation}
G(x) = \frac{A+B\,x'(x)^{-\delta}}{x'(x)^\eta}\;,
\end{equation}
that uses the variable $x'(x)$ (that cancels most of the boundary effects) and includes a short distance correcting term $Bx^{-(\eta+\delta)}$ to the large distance power law decay $A x^{-\eta}$.
In Fig.~\ref{Fig:corr_s175_scaled} we plot $G(x)\,x'(x)^\eta$ versus the correcting term $x'(x)^{-\delta}$ and we observe a rather good linear behavior (the straight line is a linear fit to the $L=2048$ data).
In the left panel we have used $\eta=0.25$ and $\delta=0.3$, while on the right panel we have used the value for $\eta$ reported by Picco in Ref.~\cite{Picco}, that is $\eta_P=0.332$, and $\delta_P=0.5$ (please notice that the results are not very sensitive to the values of $\delta$ and $\delta_P$).

By looking to the data in Fig.~\ref{Fig:corr_s175_scaled} we make two observations. Firstly, the linearity of the data in both panels is similar, with a small preference to exponents used in the left panel: so we see no reason to use an $\eta$ exponent different from the theoretical prediction, $\eta=2-\sigma$. Secondly, a more careful inspection of the data reveals that using the exponent $\eta_P$ the data at large distances (small values of $x^{-\delta}$) always tend to bend up. A possible explanation for this observation is that the value $\eta_P$ for the exponent found by Picco is somehow a compromise between the asymptotic decay $x^{-\eta}$ and the pre-asymptotic correcting term $x^{-\eta-\delta}$: indeed data in the right panel of Fig.~\ref{Fig:corr_s175_scaled} show smaller corrections (they are flatter), but eventually tend to increase because the exponent $\eta_P$ is likely to be larger than the true exponent $\eta$.

\begin{figure}[ht]
\begin{center}	
\includegraphics[width=0.8\columnwidth]{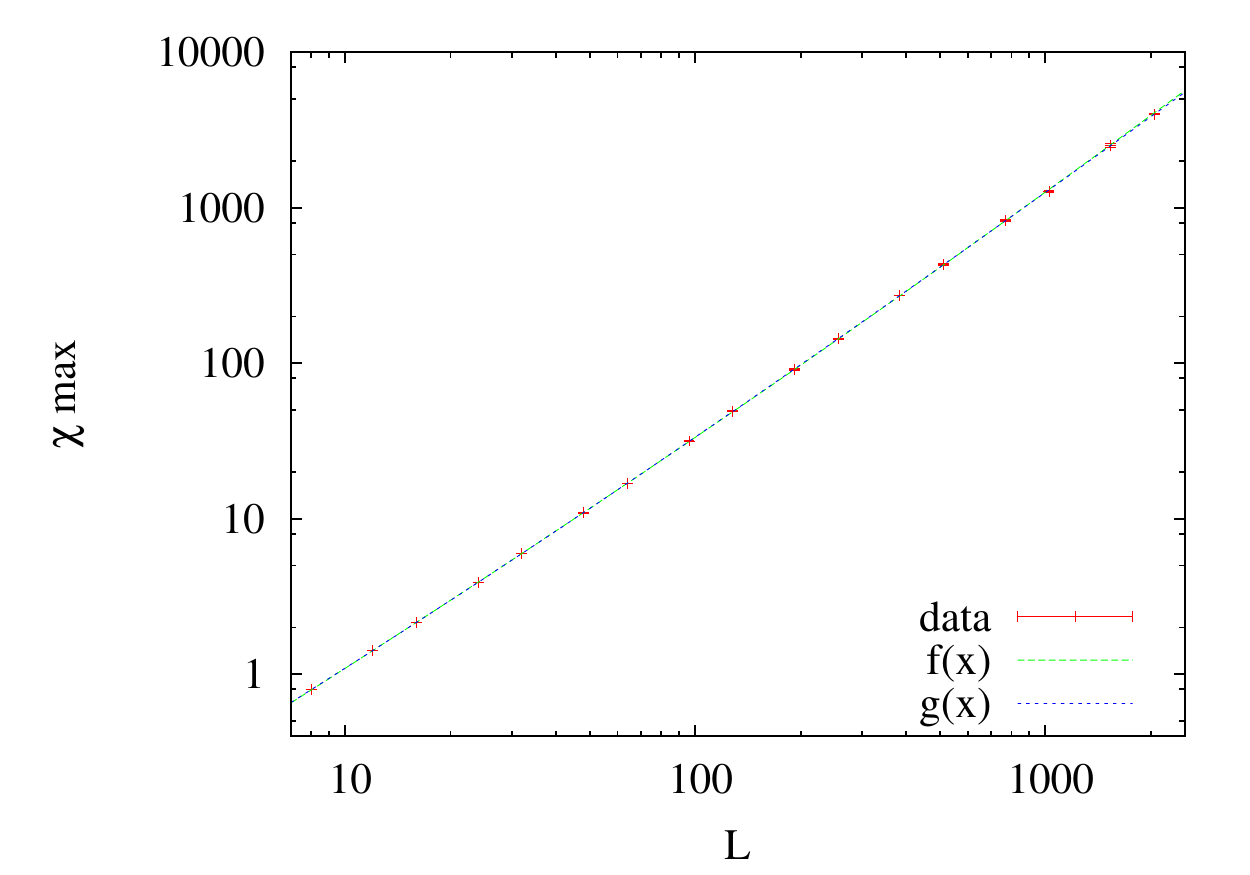}
\end{center}
\caption{\label{Fig:chifit} Log-log plot of the susceptibility at the maximum as a function of the size, for $\sigma=1.75$ and $d=2$. 
Two fits using $f(x)=L^{2-\eta}(a+bL^{-\delta})$ and $g(x)=L^{2-\eta_P}(a+bL^{-\delta_P})$ are shown, which are both compatible with the data.}
\end{figure}

The effect of the two power-laws in $G(x)$ reflects also in the measure of the $\eta$ exponent from the susceptibility. 
Indeed the susceptibility is the integral over $x$ of $G(x)$. 
This means that if we measure $\chi$ as a function of the size of the system, 
it will not follow a simple power law with exponent $2-\eta$.
Instead, it will be of the form:
\begin{equation}
\<\chi(L,T)\>=L^{2-\eta}\((a+bL^{-\delta}\))\[[F_{\chi}(L^{1/\nu}(T-T_c))+L^{-\omega}G_{\chi}(L^{1/\nu}(T-T_c))+\dots\]].
\label{Eq:FSSchi}
\end{equation}
The contribution proportional to $L^{2-\eta-\delta}$ is a new correcting term to the asymptotic behaviour, which is much bigger than the usual $L^{-\omega}$ correcting term.
Indeed, the correcting term $L^{-\omega}$ takes into account 
the fact that the correlation function saturates and stops decaying at distances close to $L/2$ 
(as can be seen in the left panel of Fig.~\ref{Fig:corr}), but at these distance $G(x)$ is small and so it is also the correcting term $L^{-\omega}$.
On the contrary, the correcting term $L^{-\delta}$ is dominant at short distances, where the correlation $G(x)$ is large and this makes the correction $L^{-\delta}$ much larger than the $L^{-\omega}$ correcting term.

If this new correcting term is not properly taken into account, then the $\eta$ value is likely to be overestimated.
This may be the reason why in Ref.~\cite{Picco} the exponent $\eta$ is found to be bigger than the one predicted by the RG analysis.
The presence of this new correcting term can be also the reason why the 
$\omega$ exponent found in the previous analysis is very small and not in agreement with the larger value of the SR model: 
actually, we think that in the correction-to-scaling analysis we are measuring $\delta$ instead of $\omega$.

For each system size we have measured the connected susceptibility at its maximum, 
which is a good proxy for the critical temperature.
Analogously to what we have done for the correlation function, 
we have performed a fit to the maximum susceptibility as a function of the size, with the sum of two power laws: either with $f(x)=L^{2-\eta}(a+bL^{-\delta})$, where $\eta=2-\sigma$, 
and with $g(x)=L^{2-\eta_P}(a+bL^{-\delta_P})$, where $\eta_P$ is the value reported by Picco in Ref.~\cite{Picco}. 
We have ignored the corrections term $L^{-\omega}$ because, as discussed above, it is much smaller than the one considered.
The results are shown in Fig. \ref{Fig:chifit}.
The values of $\delta$ obtained are $\delta=0.41$ and $\delta_P=0.43$.
The values are similar to the ones obtained from the correlation function.
Again, both scenarios are compatible with the data and much larger sizes are needed to exclude one of the two.

The two power laws behaviour is a very strange feature of the correlation function, 
because it is not present in the usual SR model, nor in the LR one far from the lower critical $\sigma$. 
We leave for a future work to understand its physical origin and to eventually provide an analytical description of it.

\subsection{Check of the superuniversality conjecture}

\begin{figure}[t]
\includegraphics[width=\columnwidth]{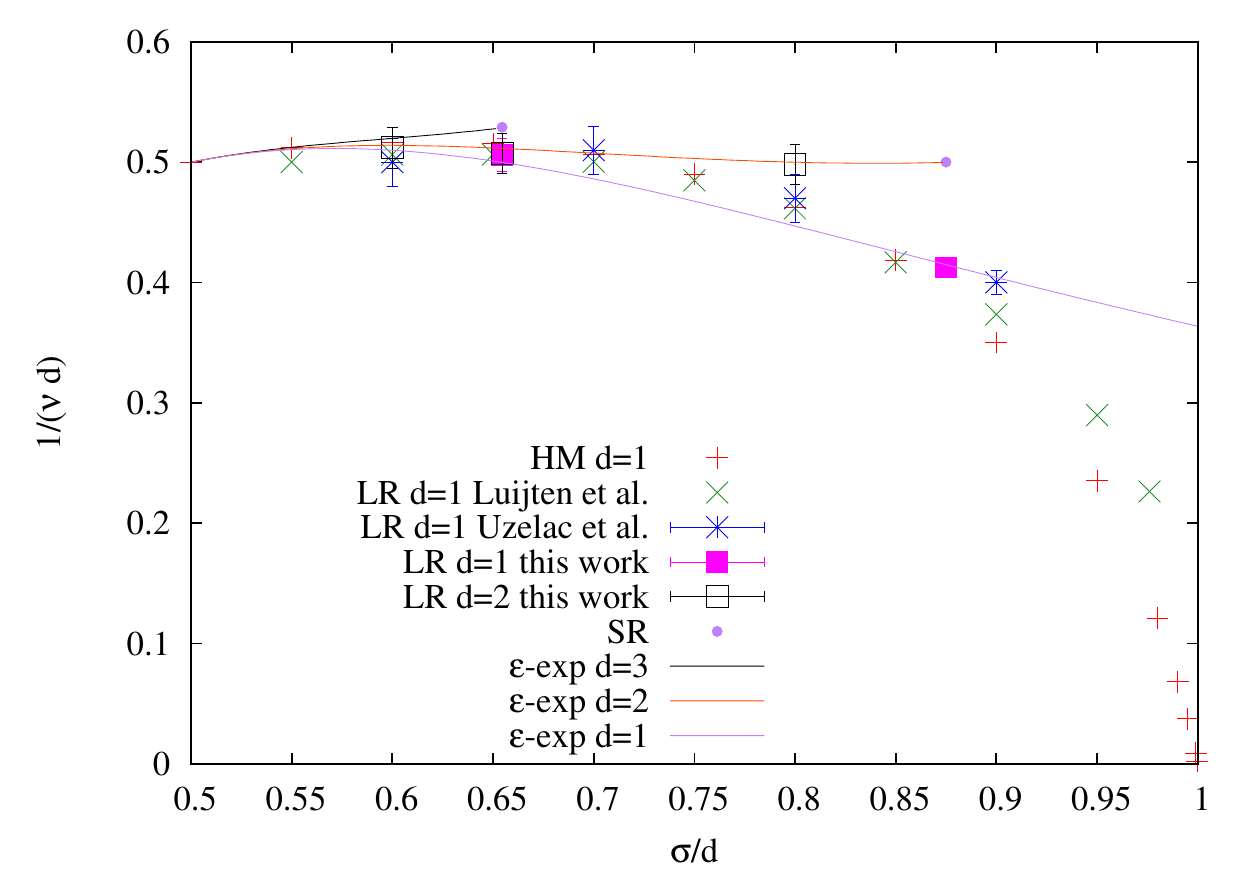}
\caption{\label{Fig:nu} $1/(\nu d)$ as a function of the exponent $\hat{\sigma}=\sigma/d$ for HM in $d=1$ and LR model in $d=1$ and $d=2$  
as found in various works and in this work. 
The SR values follow the matching formula (\ref{Eq:sigmaDd}).}
\end{figure}

At this point we want to verify the superuniversality conjecture or, equivalently, Eq. (\ref{Eq:sigmaDd}) and (\ref{Eq:nu-relation}).
For this reason we summarize the results for the critical exponents in the literature and in this work.
In Fig. \ref{Fig:nu}, $1/(d\nu)$ is plotted as a function of the parameter $\hat{\sigma}$ in the non mean-field region, 
for the HM model as found in Ref.~\cite{nu_hier}, and for the LR one-dimensional model, from Ref.~\cite{nu_LR} and \cite{nu_LR2}. 
For the LR $d=1$ and $d=2$ model, our results are reported too.
From Fig. \ref{Fig:nu} it is clear that the two analyzed one-dimensional models (HM and the LR one) 
are not in the same universality class. 
While their critical exponents are quite similar near to the upper critical $\sigma_U=1/2$, 
the differences grow approaching the lower critical $\sigma_L=1$. 
This is reasonable, because we know that the two models have very different behaviours at $\sigma_L=1$. 

To verify the exactness of Eq. (\ref{Eq:nu-relation}), 
in Fig. \ref{Fig:nu} the values of the exponent of the SR model as found in Ref.~\cite{Pellissetto} 
are placed at the corresponding value of $\hat{\sigma}$ as in Eq. (\ref{Eq:sigmaDd}): 
$\nu_{SR}(2)=1$ for $D=2$ corresponds to $\hat{\sigma}=0.875$, $\nu_{SR}(3)=0.6301(4)$ 
for $D=3$ corresponds to $\hat{\sigma}=0.65453$. 
Eq. (\ref{Eq:nu-relation}) 
is a good approximation near to the upper critical dimension (it is good for $D=3$) but it is no more good for $D=2$. 
Remembering also the results for the $\omega$ exponent, we can assert 
that it is not possible to find a single value of $\sigma$
that verify the equivalence for all the critical exponents as in Eq. (\ref{Eq:Dd-sigma_exponents}).

The lines are the third order $\epsilon$-expansion for $d=2$ and $d=3$ as found in Ref.~\cite{fieldtheoryLR}, 
where the third order term has been fixed imposing that the curves recover the SR value at $\hat{\sigma}_L(d)$, 
and the second order $\epsilon$-expansion for $d=1$. 
For $d=1$ we have not fixed the third order because at $\sigma_L(1)=1$ there is not a second order phase transition.
For this reason the curve for $1/\nu$ as a function of $\sigma$ does not approach the point $\sigma_L(1)=1$ smoothly, but with a divergent derivative.
Our data for $d=2$ are in agreement with the $\epsilon$-expansion.

\begin{figure}[t]
\includegraphics[width=\columnwidth]{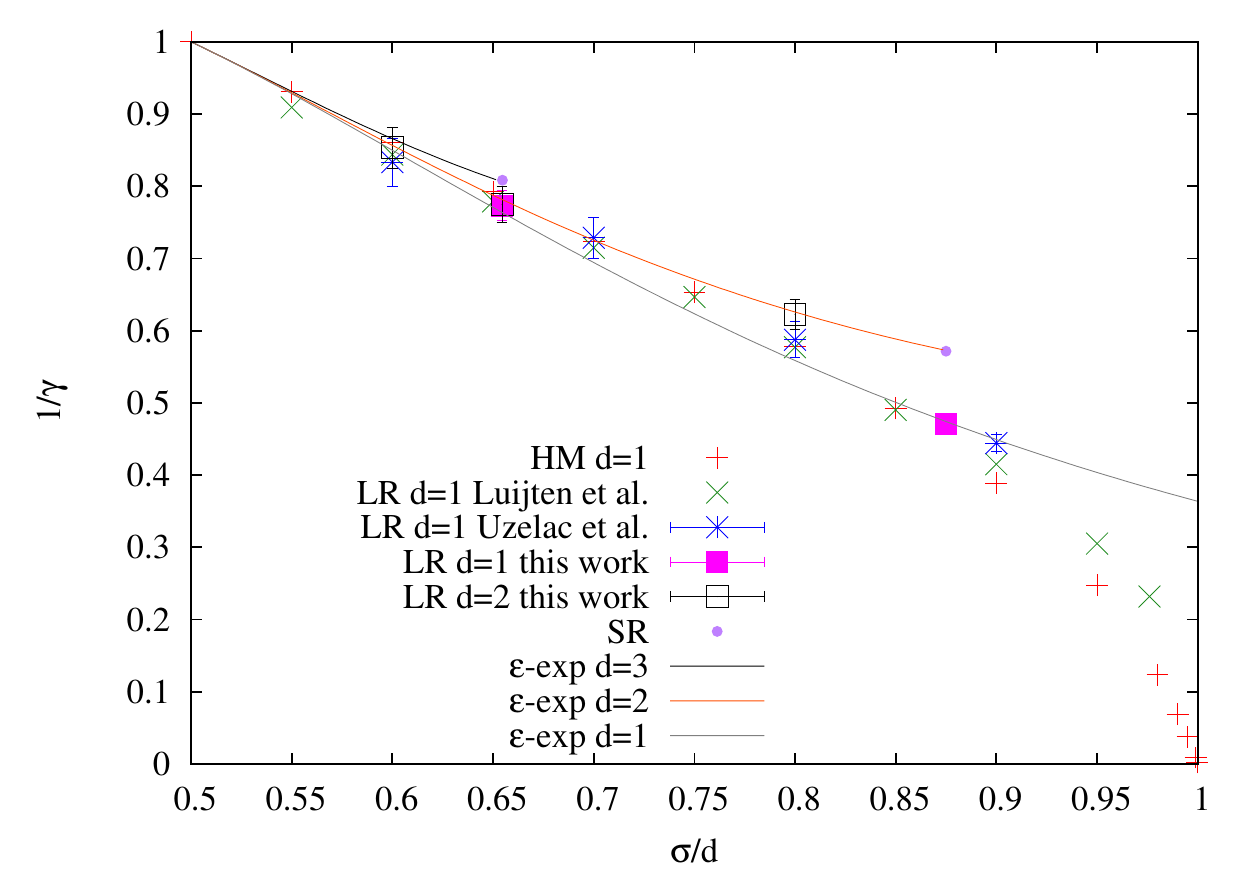}
\caption{\label{gamma} $1/\gamma$ as a function of the exponent $\hat{\sigma}=\sigma/d$ 
in the non mean-field region for HM in $d=1$ and LR model in $d=1$ and $d=2$ as found in various works and in this work.}
\end{figure}

In Fig. \ref{gamma}, $1/\gamma$ is plotted as a function of $\hat{\sigma}$ in the non mean-field region. 
The super-universality conjecture is not exact but it is a good approximation near $\sigma_U$. 
In fact $1/\gamma$ for $d=1$ and $d=2$ 
is nearly independent from $d$ and the two curves are near when plotted versus $\hat{\sigma}$.
The values for the SR model should be the end point of the line for $\gamma(\hat{\sigma})$ with $d=2$ and $d=3$, placed at $\hat{\sigma}_L(2)$ and $\hat{\sigma}_L(3)$. 
The lines are the third order epsilon-expansion as found in Ref.~\cite{fieldtheoryLR}, 
where the third order has been fixed (as before) imposing that the curves recover the SR value at $\hat{\sigma}_L(d)$.

\section{Conclusions} 

We have analyzed the connection between LR and SR systems. 
For simplicity we have considered the ferromagnetic version of the models, given that the connection 
we are interested in is still not well understood even in this simple case.
First of all we have analyzed the $d=1$ LR ferromagnetic model, 
for which the couplings have a power-law decaying with exponent $\sigma$,
and we have compared it with a SR system in $D$ dimensions.
We have reviewed all the \sDr relations proposed in the literature and we have analyzed their accuracy
performing Monte Carlo simulations to measure the exponents of the LR model through finite size scaling.
We have compared them with the exponents of SR systems available in the literature.
We have found that near to the upper critical dimension a reliable \sDr relation exists:
it means that, for example, for $D=3$ a value of $\sigma$ exists for which all the exponents of the LR and SR models are very close, 
while near to the lower critical dimension,
for example for $D=2$, it is not possible to find a value of $\sigma$ for which all the exponents of the LR model corresponds
to those of the SR one.

Then we have generalized the \sDr relation for LR systems in $d$ dimensions showing that the dimensions $D$ 
of the SR system and $d$ of the LR one enter only through their ratio $d/D$.
The \sDr relations in Eq. (\ref{Eq:Dd-sigma_exponents}) can be deduced also from a superuniversality conjecture. 
We have verified this property performing Monte Carlo simulations at various values of $\sigma$ for $d=2$ 
to measure the critical exponents.
The superuniversality conjecture is a good approximation near to the upper critical dimension and
becomes worst going towards the lower critical one.

Finally for the $d=2$ LR model we have studied the region near to the lower critical dimension $\sigma_L=1.75$.
We have discovered that in this region the correlation function has a very strange behaviour, 
characterized by two decaying power-laws.
This makes difficult to measure with high precision the critical exponents and the lower critical dimension.
Standard finite size scaling arguments do not help since the subdominant power law has an effect much larger than leading order finite size effects.
This kind of critical correlation function (with two different power laws) 
can easily lead to overestimate $\eta$ if a proper fit with a double power-law is not performed.
Although we have performed the improved fit with two power laws, the exponent of the asymptotic decay, 
i.e.\ the critical exponent $\eta$, has a very large uncertainty, 
that makes it compatible both with the standard RG calculation by Sak \cite{Sak} and with the recent proposal by Picco \cite{Picco}.
According to Occam's razor, we see no reason to propose a different scenario \cite{Picco,Picco2} 
as long as the numerical data, properly fitted, are compatible with the standard RG scenario proposed by Sak 40 years ago \cite{Sak}.

An important theoretical challenge is to understand analytically the origin of the two power laws 
appearing in the critical correlation function close to the lower critical $\sigma$.

\begin{acknowledgments}
This research has received financial support from the European
Research Council (ERC) through grant agreement No. 247328 and from the
Italian Research Minister through the FIRB project No. RBFR086NN1.
\end{acknowledgments}

\end{document}